# Vortex-induced Shear Polaritons

*Shuwen Xue*[1†], *Yali Zeng*[1†], *Sicen Tao*[1], *Tao Hou*[1], *Shan Zhu*[1], *Chuanjie Hu*[1],

*Huanyang Chen*[1,2]

1.Institute of Electromagnetics and Acoustics and Department of Physics, College of Physical Science and Technology, Xiamen University, Xiamen 361005, China

2. e-mail: kenyon@xmu.edu.cn

†These authors contributed equally to this Letter

**ABSTRACT:** Hyperbolic shear polaritons (HShPs) emerge with widespread attention as a new class of polariton modes with broken symmetry due to shear lattices. In this letter, we find a new mechanism of generating HShPs. When utilizing vortex waves as excitation sources of hyperbolic materials without off-diagonal elements, HShPs will appear. In addition, this asymmetric HShPs can be recovered as symmetric modes away from the source, with a critical transition mode between the left-skewed and right-skewed HShPs, via tuning the magnitude of the off-diagonal imaginary component and controlling the topological charge of vortex source. It is worth mentioning that we explore the influence of parity of topological charges on the field distribution and demonstrate these exotic phenomena from numerical and analytical perspectives. Our results will promote new opportunities for both HShPs and vortex waves, widening the horizon for various hyperbolic materials based on vortex sources and offering a new degree of freedom to control various kinds of polaritons.

**INTRODUCTION**

Polaritons are half-light and half-matter quasiparticles formed by the interaction between light and matter modes involving collective oscillations of polarization charges in matter, enabling nanoscale control of light. Traditional surface plasmon polaritons (SPPs) are electromagnetic waves that travel along the metal–dielectric interface, which have received a lot of attention[1-4]. In addition, exhibiting long lifetimes and low optical losses, phonon polaritons (PhPs) emerge as an important substitute for plasmonic counterparts in sub-diffraction light-matter interaction and nanophotonic applications, such as the PhPs in polar van der Waals (vdWs) materials including hexagonal boron nitride (hBN)[5-7], α-phase molybdenum oxide (α-$MoO_3$) [8-11] and so on. Remarkably, α-$MoO_3$ with capable of supporting topological transitions enables both in-plane hyperbolic and elliptical PhPs[12]. Furthermore, a new polariton class in low-symmetry monoclinic and triclinic crystals (e.g., β-phase $Ga_2O_3$) called hyperbolic shear polaritons (HShPs) complements the previous observations of hyperbolic PhPs and exists when the dielectric tensor is not diagonalized[13]. This drives us to consider the possibility of the generation of HShPs in orthogonal systems (with three major polarizability axes) without off-diagonal elements.

As mentioned earlier, PhPs with long lifetimes and low optical losses offer more opportunities for infrared nanophotonic applications. Yet, most methods of stimulating

PhPs are based on simple dipole source, and it is high time to take complex structured fields into account. Vortex as a representative complex structured field, not only exists in nature, such as spiral galaxies in Milky Way and typhoon vortices, but also is extensively applied in structured electromagnetic and optical fields. Optical vortex is a beam of photons that propagates with singularity on its axis in the form of $e^{im\theta}$, where $m$ and $\theta$ represent the topological charge and azimuth angle, respectively. Compared with the conventional plane wave or dipole source, vortex beams with a higher degree of freedom attract enormous applications and interest, including vortex tweezers [14-15], high-capacity optical communications[16], optical microscopy imaging[17-18], nonlinear optics[19-20] and so on. Nonetheless, most of these vortex-induced phenomena are confined to isotropic materials and it is of great significance to study the propagation characteristics of vortex in anisotropic materials (or even hyperbolic materials), which remains unexplored.

In this letter, based on the motivations mentioned above, we propose an approach to induce HShPs in orthogonal systems without off-diagonal elements and study the propagation characteristics of vortex waves as excitation sources in hyperbolic materials. In details, asymmetric HShPs will occur when the vortex waves are used as excitation sources of hyperbolic materials, providing a new degree of freedom to control various polaritons. More interestingly, asymmetric HShPs (left-skewed and right-skewed HShPs) excited by the vortex source can be recovered as symmetric modes by regulating the magnitude of the off-diagonal imaginary component, arising the critical symmetry transition. We verify these unique phenomena theoretically and numerically, providing a platform for future research of HShPs and vortex waves.

**RESULTS AND DISCUSSION**

The diagram in Fig. 1 illustrates the simulated magnetic fields ($H_z$), the corresponding intensities ($|H|$) and fast Fourier transform (FFT ($\text{Re}[H_z]$)) of vortex wave with different topological charges ($m = 0, \pm 1$) as excitation source of hyperbolic materials at 718 cm$^{-1}$, where the simulated permittivity tensor is obtained by diagonalizing the real part of the permittivity tensor of β-phase Ga$_2$O$_3$ ($\text{Re}[\varepsilon(\omega)]$) individually at each frequency. Here, the off-diagonal permittivity tensor elements are set as zero and the corresponding material parameters ($\varepsilon_{mm}$ and $\varepsilon_{nn}$) are achieved from Ref [13]. For the conventional point source ($m = 0$), the simulated field expresses hyperbolic feature and symmetry apparently as shown in Fig. 1(b). On the contrary, if utilizing vortex wave as excitation sources ($m = \pm 1$), the simulated magnetic field tips to one side (Fig. 1(a) and (c)), presenting asymmetric phenomenon obviously. For the opposite topological charges $m = \pm 1$, the simulated magnetic fields tips to the left side if $m = -1$ while tips to the right side if $m = +1$. The corresponding intensities ($|H|$) and FFT dispersions are displayed in Fig. 1(d)-(i), which tips with opposite directions for opposite topological charges as we predicted before. Therefore, the asymmetric HShPs appears when vortex waves serve as excitation sources of hyperbolic materials, validating the possibility of generating asymmetric HShPs in orthogonal systems without off-diagonal elements. The simulated fields were obtained with finite element software COMSOL Multiphysics.

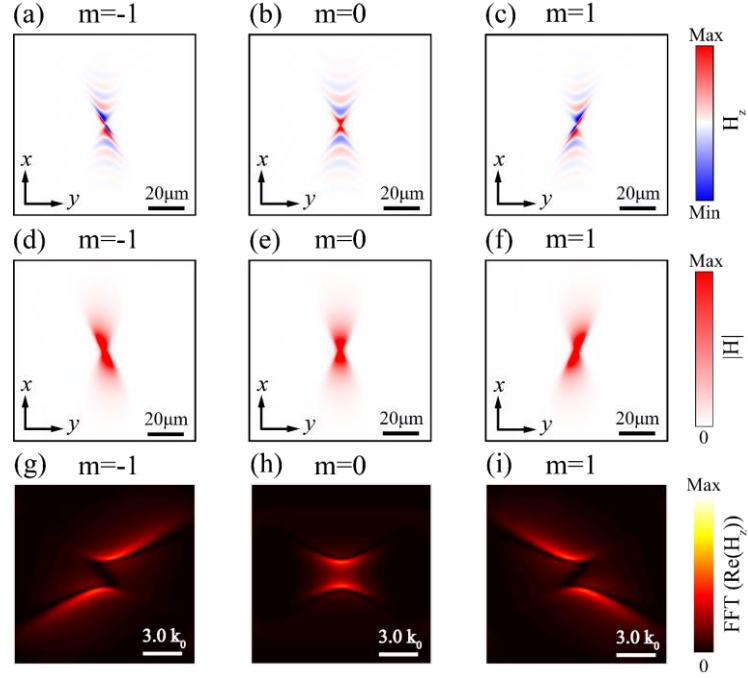

**Figure 1.** HShPs based on the vortex waves with different topological charges ( $m = 0, \pm 1$ ) as excitation sources of hyperbolic materials without off-diagonal permittivity tensor elements at 718 cm$^{-1}$. (a)-(c) are the simulated magnetic fields of different topological charges. (d)-(f) are the corresponding intensities ($|H|$) of different topological charges. (g)-(i) are the corresponding fast Fourier transform (FFT ( $\text{Re}[H_z]$ )) of different topological charges.

In order to further demonstrate the phenomena of Fig. 1, we shall prove in more detail from an analytical perspective. Above all, let us focus on the 2D transverse magnetic (TM) polarization, and Maxwell's equations can be described as

$$\begin{cases} \nabla \times \vec{E} = i\omega\mu_0 \hat{\mu}\vec{H} \\ \nabla \times \vec{H} = -i\omega\varepsilon_0 \hat{\varepsilon}\vec{E} \end{cases} \quad (1).$$

For the point source in the free space, the general solution can be expressed as $H_{\text{point}} = H_m^{(1)}(k_0 r)e^{im\theta}$, where $H_m^{(1)}$ is the $m$th-order Hankel functions of the first kind, $r = \sqrt{x^2 + y^2}$, $\theta = \arccos(x/r)$ for $y > 0$ and $\theta = 2\pi - \arccos(x/r)$ for $y < 0$. And $m = 0$ is the order of a point source. Likewise, the general solution is appropriate for the vortex waves as excitation sources and $m$ is set as other non-zero integer values. Considering the anisotropic case, a mapping transformed into an isotropic space can be defined as

$$x = \sqrt{\varepsilon_{nn}} x', \; y = \sqrt{\varepsilon_{mm}} y' \tag{2}$$

where $\varepsilon_{mm}$ and $\varepsilon_{nn}$ are the permittivity tensors in a rotated coordinate system [$mnz$]. Therefore, $r, \theta$ and the general solution of hyperbolic materials in ($x'$, $y'$) coordinate could be rewritten as $r' = \sqrt{\varepsilon_{nn} x'^2 + \varepsilon_{mm} y'^2}$, $\theta' = \arccos(\sqrt{\varepsilon_{nn}} x'/r')$ if $y > 0$, $\theta' = 2\pi - \arccos(\sqrt{\varepsilon_{nn}} x'/r')$ if $y < 0$ and $H_{vortex} = H_m^{(1)}(k_0 r') e^{im\theta'}$, respectively. Through the derivations, the analytical results excited by different topological charges ($m = \pm 1, \pm 2, \pm 3, \pm 4$) in anisotropic materials at 718 cm$^{-1}$ can be determined, which is appropriate for both cases of hyperbolic and elliptical materials. As shown in Fig. 2, the magnetic fields present the right-skewed HShPs for positive topological charges, while negative topological charges stimulate the left-skewed HShPs. The asymmetric effect becomes more remarkable as the topological charge increases. Additionally, for odd topological charges, the phase of the upper half differs $\pi$ from that of the lower half, corresponding to the blue and red part of $m = \pm 1, \pm 3$. In contrast, the phase of the upper half is the same as that of the lower half for even topological charges ($m = \pm 2, \pm 4$), whose magnetic fields present a rotational symmetry distribution and can transform each other (the upper half and lower half) by rotating 180° around the origin. Notice that the distribution of the magnetic field near the source, whose patterns are like needles, is associated with the topological charges and changes regularly as the topological charge changes. Our analytical results in Fig. 2 prove a high level of consistency and correctness of the numerical results in Fig. 1. Up to now, the new mechanism of generating HShPs by utilizing vortex waves as excitation source has been demonstrated, which is based on the hyperbolic materials without off-diagonal elements and is different from the previous HShPs based on the off-diagonal elements of permittivity tensors.

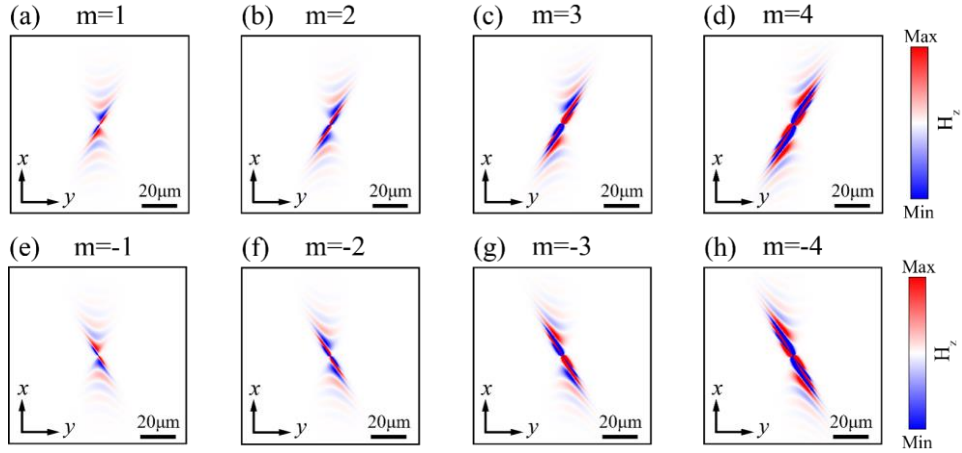

**Figure 2.** Analytical results of HShPs based on the vortex with different topological charges ($m = \pm 1, \pm 2, \pm 3, \pm 4$) as excitation source of hyperbolic materials without off-diagonal permittivity tensors at 718 cm$^{-1}$. (a)-(d) are the analytical magnetic fields of positive topological charges. (e)-(h) are the analytical magnetic fields of negative topological charges.

From the derivation and discussion above, we know that the general solution of vortex waves in anisotropic materials could be denoted as $H_{vortex} = H_m^{(1)}(k_0 r')e^{im\theta'}$ with $\theta' = \arccos(\sqrt{\varepsilon_{nn}} x'/r)$ for $y > 0$ and $\theta' = 2\pi - \arccos(\sqrt{\varepsilon_{nn}} x'/r)$ for $y < 0$. To better understand the asymmetric phenomenon, we plot the imaginary component of $\theta'$ for $\varepsilon_{nn} > 0$ and $\varepsilon_{nn} < 0$ in Fig. 3, respectively, and select $\varepsilon_{mm} = 1 - 0.3i, \varepsilon_{nn} = -3 - 0.3i$ and $\varepsilon_{mm} = 1 - 0.3i, \varepsilon_{nn} = 3 - 0.3i$ as contrast examples. Since the existence of negative electric effective tensor elements of hyperbolic materials, $\theta'$ has the imaginary component as shown in Fig. 3(a) and $e^{im\theta'}$ becomes a complex number with real and imaginary component. Therefore, for positive imaginary component (red regions of Fig. 3(a)), $e^{im\theta'}$ becomes an attenuation factor while $e^{im\theta'}$ becomes a gain factor if imaginary component is negative (blue region), causing the asymmetric HShPs. On the contrary, if $\varepsilon_{nn} > 0$, the imaginary component of $\theta'$ is near to zero (Fig. 3(d)) and the modulus of $e^{im\theta'}$ is identically equal to 1. Based on the analysis above, we plot the magnetic fields and intensities ($|H|$) of $\varepsilon_{nn} < 0$ and $\varepsilon_{nn} > 0$ in Fig. 3(b)-(c) and (e)-(f), respectively. As we have predicted, for the hyperbolic case ($\varepsilon_{nn} < 0$), the magnetic fields and intensities demonstrate the asymmetric phenomenon (Fig. 3(b)-(c)) obviously while become symmetric for the elliptical case ($\varepsilon_{nn} > 0$) (Fig. 3(e)-(f)). These precise and straight analysis not only produce evidence for the generation of asymmetric HShPs in hyperbolic materials without off-diagonal elements, but also provide the guidance for future exploration of hyperbolic materials.

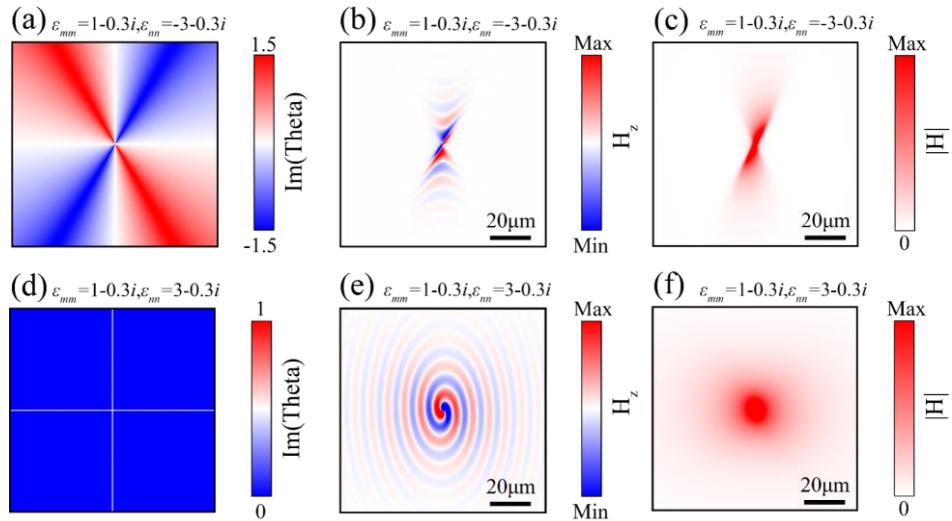

**Figure 3.** Explanation of asymmetric HShPs in hyperbolic materials. (a) and (d) are the imaginary components of $\theta'$ for $\varepsilon_{mm} = 1-0.3i, \varepsilon_{nn} = -3-0.3i$ and $\varepsilon_{mm} = 1-0.3i, \varepsilon_{nn} = 3-0.3i$, respectively.

(b) and (e) are the corresponding magnetic fields of vortex waves ($m=1$) as excitation sources at 718 cm$^{-1}$. (c) and (f) are the corresponding intensities of vortex waves ($m=1$) as excitation sources at 718 cm$^{-1}$.

After demonstrating the generation of HShPs in orthogonal systems without off-diagonal elements, we intend to introduce the off-diagonal imaginary component ($\text{Im}(\varepsilon_{mn})$) and explore the propagation characteristics of polaritons excited by vortex sources, as illustrated in Fig. 4, with the permittivity tensors expressed as

$$\begin{pmatrix} \varepsilon_{mm} & i \times f\, \text{Im}(\varepsilon_{mn}) & 0 \\ i \times f\, \text{Im}(\varepsilon_{mn}) & \varepsilon_{nn} & 0 \\ 0 & 0 & \varepsilon_{zz} \end{pmatrix} \quad (3)$$

where $f$ is the scaling factor for tuning the magnitude of the off-diagonal imaginary component. To measure the asymmetry of the field distributions, we first integrate the modulus of the magnetic field along the dashed red line and blue line ($x = 25\,\mu m$) in Fig. 4(b) at 718 cm$^{-1}$, respectively, and then calculate the difference between the obtained integral values (the left-hand side minus the right-hand side). The corresponding results for different topological charges m and scaling factors $f$ (corresponding to the horizontal and vertical coordinates) are displayed in Fig. 4(a). The red region in Fig. 4(a) represents the left-skewed HShPs that the intensity of polaritons propagating along the left stronger than that of the right, while the blue region denotes the right-skewed HShPs. The black dashed lines in Fig. 4(a) indicate the critical symmetry transition between the left-skewed HShPs and right-skewed HShPs. It is obvious that the symmetry of magnetic fields away from the source can be regulated through tuning the topological charges m and scaling factors $f$, undergoing the transition of left-skewed HShPs to symmetric HShPs, and then to right-skewed HShPs. Concretely, the symmetric magnetic fields (away from the source) for different topological charges ( $m = 0, \pm1, \pm2$ ) and scaling factors are illustrated in Fig. 4(b)-(f). As the topological charge m increases, the asymmetrical effect also increases and a larger scaling factor f is needed to return to symmetry. Here a positive scaling factor is required for a positive topological charge to recover as symmetry and vice versa. Compared with the previous generation of HShPs by tuning the scaling factors *f*, we propose a new mechanism of generating HShPs and introduce a new degree of freedom (topological charges m) to control the symmetry (away from the source), which not only possesses the restorability and tunability, but also provides new opportunities for both HShPs and vortex waves to some extends.

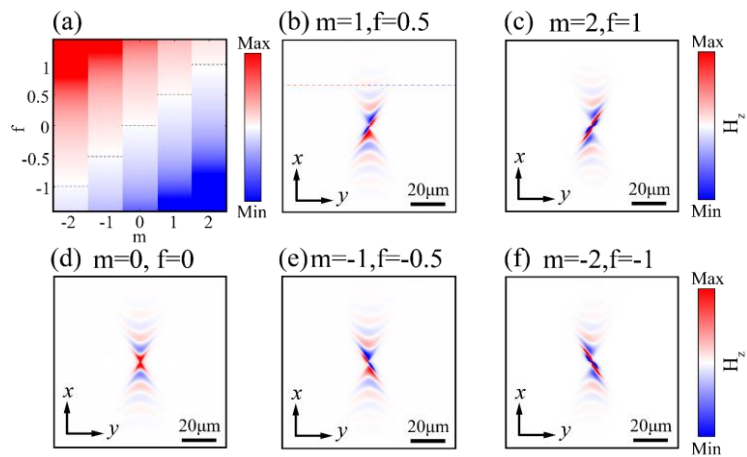

**Figure 4.** The critical symmetry transition between the left-skewed and right-skewed HShPs induced by the vortex waves at 718 cm$^{-1}$. (a) The symmetry transition for different topological charges and different scaling factors. (b)-(f) The corresponding symmetric magnetic fields (away from the source) for different topological charges ( $m = 0, \pm1, \pm2$ ) and different scaling factors ( $f = 0, \pm0.5, \pm1$ ) at 718 cm$^{-1}$, respectively.

To demonstrate the whole transition process, we choose the topological charge $m=1$ as a specific example in Fig. 5. Compared with the symmetric magnetic field (away from the source) in Fig. 5(b) ($m=1, f=0.5$), both decreasing and increasing the scaling factor f, the asymmetric effect of HShPs will be amplified, exhibiting more distinct left-skewed HShPs and right-skewed HShPs, as shown in Fig. 5(a) and (c) ($f=1.5$ and $f=-0.5$). Being consistent with the distribution of Fig. 4(a), when raising the scaling factor $f$, the magnetic fields away from the source will tip to the left side (Fig. 5(a)), corresponding to the red region in Fig. 4(a). And the magnetic fields away from the source will tip to the right side if decreasing the scaling factor $f$ (Fig. 5(c)), corresponding to the blue region in Fig. 4(a). The corresponding intensities ($|H|$) are displayed in Fig. 5(d)-(f), which present the whole transition process directly.

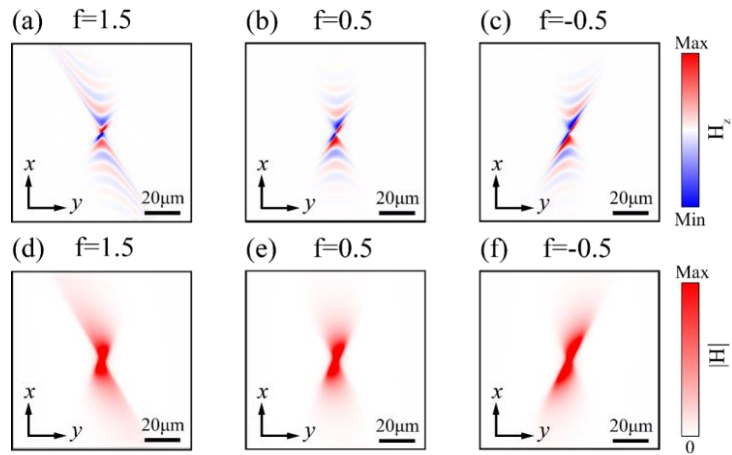

**Figure 5.** More distinct asymmetric HShPs for decreasing and increasing scaling factors $f$ induced by the vortex waves ($m=1$) at 718 cm$^{-1}$. (a)-(c) are the magnetic fields of different scaling factors ($f=+1.5, \pm 0.5$). (d)-(f) are the corresponding intensities ($|H|$) of different scaling factors ($f=+1.5, \pm 0.5$).

**CONCLUSION**

In summary, we have proposed a new approach to generate HShPs and demonstrated its feasibility from both numerical and analytical perspectives. More specifically, if the vortex source is introduced into hyperbolic materials, the HShPs will appear in materials without off-diagonal elements and its magnetic fields are related to the parity of topological charges. Moreover, through governing the magnitude of the off-diagonal imaginary components and topological charges, the symmetry of magnetic fields away from the source can be tuned, which offers a restorable and tunable way for future research of HShPs. It should be noted that our work provides a new degree of freedom for tuning the HShPs and vortex waves, which has tremendous promise for implementation and expansion in other sources, such as Janus and Huygens dipoles[21].

**AUTHOR INFORMATION**

**Corresponding Authors**

**Huanyang Chen** – Department of Physics, Xiamen University, Xiamen, 361005, P. R. China; Email: kenyon@xmu.edu.cn

**Authors**

**Shuwen Xue** – Institute of Electromagnetics and Acoustics, Xiamen University, Xiamen, 361005, P. R. China

**Yali Zeng** – Department of Physics, Xiamen University, Xiamen, 361005, P. R. China

**Sicen Tao** – Institute of Electromagnetics and Acoustics, Xiamen University, Xiamen, 361005, P. R. China

**Tao Hou** –Institute of Electromagnetics and Acoustics, Xiamen University, Xiamen, 361005, P. R. China

**Shan Zhu** –Department of Physics, Xiamen University, Xiamen, 361005, P. R. China

**Chuanjie Hu** –Institute of Electromagnetics and Acoustics, Xiamen University, Xiamen, 361005, P. R. China

**Author Contributions**

[†] S.X. and Y.Z. contributed equally to this work.

**Notes**

The authors declare no competing financial interest.

**ACKNOWLEDGMENT**

This work is supported by the National Natural Science Foundation of China (Grants No. 11874311, 92050102), the National Key Research and Development Program of China (Grant No. 2020YFA0710100), Fundamental Research Funds for the Central Universities (Grants No. 202006310051, 20720220134, 20720220033).